\documentclass[epj]{webofc}
\usepackage[utf8]{inputenc}
\usepackage[varg]{txfonts}   
\usepackage{booktabs}
\usepackage{xcolor}
\definecolor{darkred}{rgb}{0.4,0.0,0.0}
\definecolor{darkgreen}{rgb}{0.0,0.4,0.0}
\definecolor{darkblue}{rgb}{0.0,0.0,0.4}
\usepackage[bookmarks,linktocpage,colorlinks,
    linkcolor = darkred,
    urlcolor  = darkblue,
    citecolor = darkgreen]{hyperref}
\usepackage{subfigure}
\usepackage{slashed}
\usepackage{hepunits}
\wocname{EPJ Web of Conferences}
\woctitle{Lattice2017}
\newcommand{\msbar}{{\overline{\rm MS}}}

\newcommand{\bare}{{\rm bare}}

\begin{document}
%
\selectlanguage{english}
\title{%
Topological susceptibility with a single light quark flavour
}
\author{%
\firstname{Julien} \lastname{Frison}\inst{1}\fnsep\thanks{Speaker, \email{jfrison@post.kek.jp}} \and
\firstname{Ryuichiro} \lastname{Kitano}\inst{1,2} \and
\firstname{Norikazu}  \lastname{Yamada}\inst{1,2}\fnsep
}
\institute{%
High Energy Accelerator Research Organization (KEK), Tsukuba 305-0801, Japan
\and
Department of Particle and Nuclear Physics, The Graduate University for Advanced Studies (Sokendai), Tsukuba 305-0801, Japan
}
\abstract{%
    One of the historical suggestions to tackle the strong CP problem is to take the up quark mass to zero while keeping $m_d$ finite. The $\theta$ angle is then supposed to become irrelevant, i.e. the topological susceptibility vanishes. However, the definition of the quark mass is scheme-dependent and identifying the $m_u=0$ point is not trivial, in particular with Wilson-like fermions. More specifically, up to our knowledge there is no theoretical argument guaranteeing that the topological susceptibility exactly vanishes when the PCAC mass does.

    We will present our recent progresses on the empirical check of this property using $N_f=1+2$ flavours of clover fermions, where the lightest fermion is tuned very close to $m^{PCAC}_u$=0 and the mass of the other two is kept of the order of magnitude of the physical $m_s$. This choice is indeed expected to amplify any unknown non-perturbative effect caused by $m_u\not=m_d$. The simulation is repeated for several $\beta$s and those results, although preliminary, give a hint about what happens in the continuum limit.
}
\maketitle
\section{Introduction}\label{sec:intro}

A standard method to derive field theories is to define some basic symmetries, and then write in the lagrangian all the terms allowed by those symmetries. In QCD, we therefore have no particular theoretical reason not to include the $\theta F\tilde{F}$ term. However, we do not have any experimental evidence of a non-zero $\theta$ parameter, which would appear as a source of CP violation in QCD. In neutron electric dipole moment (nEDM) for instance, everything happens as if $\theta$ was exactly zero, or at most $O(10^{-10})$. The lack of explanation for this empirical observation is called the strong CP problem. 

One popular solution to this problem is the Peccei-Quinn mechanism, in which a new particle called axion dynamically sets the effective $\theta$ to zero. This axion is then also a dark matter candidate and is constrained by experiments. 

On the other hand, a simpler scenario has been proposed in which no axion is needed and the lack of CP-violating observables comes from QCD alone. Indeed, a non-zero $\theta$ term can be absorbed by the mass term as a complex phase, and if one flavour (the up quark) happens to be massless then it seems that the $\theta$ term is physically irrelevant since $m_ue^{i\theta}$ is always zero. In particular, the topological susceptibility (along with any higher moment of the topological charge, i.e. any contribution of configurations with non-trivial topology) should disappear, since it is a physical observable encoding the $\theta$ dependence of the free energy.

\section{Renormalisation pattern}\label{sec:renorm-pattern}

  For a general choice of regulator and renormalisation scheme, a mass of flavour $f$ will get renormalised as
  \begin{equation}
    m_f^{ren} = Z_f(\beta,m_{f'}) m_f^{bare} + m_f'(\beta,m_{f'}) .
  \end{equation}

  Massless renormalisation schemes form a specific class of schemes which allow a multiplicative renormalisation
  \begin{equation}
    m_f^{ren} = Z(\beta) \left[m_f^{bare}-m_{crit}(\beta)\right] ,
  \end{equation}
  which is convenient to give renormalised mass ratios. At very high energies any scheme will converge to a scheme of this class, with the perturbative leading terms being universal. 

  And another class of renormalisation scheme is made of those compatible with the PCAC relations
  \begin{equation}
    \partial A_\mathrm{non-singlet} = 2mP_\mathrm{non-singlet} \quad\mathrm{or}\quad\partial A_\mathrm{singlet} = 2mP_\mathrm{singlet} - \frac{1}{32\pi^2} F\tilde F .
  \end{equation}
  In the RI/MOM scheme for instance the renormalisation factors are compatible with the PCAC relation at high energy, although both explicit and spontaneous chiral breaking introduce additional non-perturbative contributions at low energy.

  Every of those schemes will correspond to a different definition of what $m_f^{ren}=0$ means. While for $N_f=2$ this ambiguity is limited by the physical interpretation of $m=0$ corresponding to a massless pion, this argument is lost when there is only one flavour of light quark. If, in particular, we want a scheme compatible with the property that the topological susceptibility vanishes when $m_u$ does, it is not obvious whether this is compatible with the axial Ward identity and multiplicative renormalisation. 

  It has in particular been suggested that a 't Hooft vertex connected to mass insertions of different flavours could bring a non-perturbative additive renormalisation to $m_u$, expected to be something like
  \begin{equation}
    \Delta m_u \sim \frac{m_dm_s}{\Lambda_{QCD}}
  \end{equation}
  for renormalisation scales $\Lambda\sim\Lambda_{QCD}$.

\section{Current knowledge on $m_u$}\label{sec:current-know}

  As of today, the FLAG~\cite{Aoki:2016frl} quotes different determinations of $m_u$ which are combined as
  \begin{equation}
    m_u^{\msbar, 2\ \GeV} = 2.16(9)(7) .
  \end{equation}

  This seems to strongly exclude the $m_u=0$ solution to the CP problem. 

  However this might just be a ``wrong'' definition of $m_u$. 
  The physical meaning of the $m_u=0$ solution to the strong CP problem is 
  actually $\chi_t = 0$, which is, in fact, not guaranteed to happen at $m_u=0$ 
  in an arbitrary scheme or scale. Therefore, it is highly nontrivial whether 
  $\chi_t=0$ requires $m_u^{\msbar}$ to be zero, because of 
  non-perturbative and scheme-dependent effects happening at low energy~\cite{Creutz:2003xc}. 

  In 2017, the Particle Data Group~\cite{Patrignani:2016xqp} still claims: 
\begin{quote}
``The estimates of d and u masses are not without controversy and remain under active investigation. Within the literature there are even suggestions that the u quark could be essentially massless.''
\end{quote}

\section{Strategy}\label{sec:strategy}

One method to tackle this issue would be to simply stop using $m_u$ as an observable constraining the $m_u=0$ solution, as paradoxical as it may sound, and instead use only $\chi_t$. To hide the paradox we could choose to rename this solution the ``axionless solution'' for instance, since an explicit knowledge of $m_u$ in an explicitly defined scheme is not needed for this solution to be tested. Therefore, any work which undertakes the tuning of $m_u$ to its physical value (e.g. by reproducing $M_\pi,M_K,M_\Sigma$) in a fully controlled way could give some reliable constraint on the ``axionless solution'' by calculating $\chi_t$ (a very cheap observable, which even comes from free if one has already monitored $Q$ to ensure a good topological tunneling). 
However, we take a more direct path to study the relation between $m_u$ and $\chi_t$. 

  In order to make more visible any effect such as $O(m_dm_s/\Lambda)$ additive renormalisation, we generate a set of $N_f=1+2$ lattices where $m_d$ is chosen unphysically large, degenerate with $m_s$. By contrast, $m_u$ is chosen such that the $m_u$ ``PCAC mass'' (determined from a combination of non-singlet Ward identities) is as close as possible to zero. This is obtained by using RHMC on the up quark, tuning its parameters so that the spectrum of $D^\dag D$ should be covered by the approximation range even for very light quark.

  We chose a L\"uscher-Weisz tree-level improved action with two steps of HEX smearing and clover fermions, in order to stay close to one of the main determinations~\cite{Fodor:2016bgu} of $m_u^{\msbar}$.

  Because $\chi_t$ is known to be non-zero at finite lattice spacing even in $N_f=2+1$, we consider the continuum limit as a crucial step and have generated configurations at several $\beta$ ($3.31,3.5,3.61,3,7,3.8$). As a first step we computed $\chi_t$ for a wide range of $m_u$ on the coarsest ensemble. Surprisingly, it turned out to be realistically computable for very low or even slightly negative $m_u^{PCAC}$, which probably corresponds to the absence of light meson when only one flavour goes to the chiral limit. We therefore focus in a second step on generating ensembles directly at near-zero $PCAC$ up mass.

  Most of the ensembles are $16^3\times 32$, which for the finer ensembles makes tiny lattices in physical units, but it is expected to be reasonable given the absence of light meson in $N_f=1+2$. A few $24^4\times48$ ensembles have been generated to check for finite-volume effects.  

  Ideally we would like to obtain the topological susceptibility through different methods, but we eventually had to settle on a gluonic definition (5-Li) combined with large time gradient flow (for different choices of flow action). 

  The PCAC masses we obtain will be computed from
  \begin{equation}
    m_u^{PCAC} = m_{ud}^{PCAC}-m_{ds}^{PCAC}/2 , 
  \end{equation}
  where $m_{ff'}^{PCAC}$ is a non singlet PCAC mass obtained from
  \begin{equation}
    m_{ff'}^{PCAC} = \frac{\partial_0 \langle (f\gamma_0\gamma_5f') (f\gamma_5f')\rangle}{\langle (f\gamma_5f') (f\gamma_5f')\rangle}  .
  \end{equation}

\section{Results}\label{sec:results}
 
We show in Table~\ref{tab:ensembles} the ensembles which have been generated for this project. We have explored a large set of parameters but could only afford small lattices. In Tables~\ref{tab:m1624}~and~\ref{tab:chi1624} we argue that the finite volume effects remain under control even for those small lattices, thanks to the absence of light pion. 

\begin{figure}[thb]
  \begin{center}
    \includegraphics[width=\linewidth]{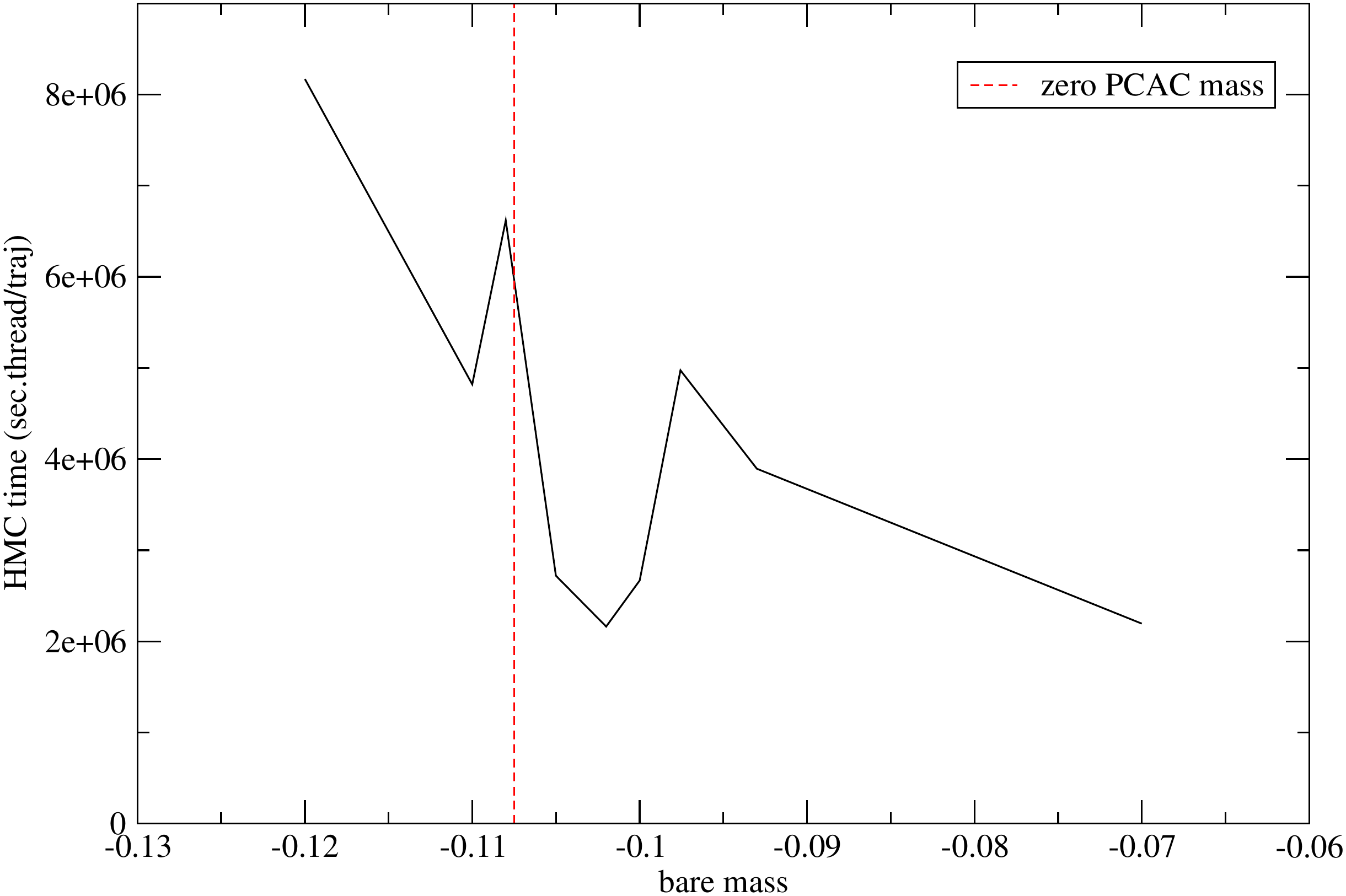}
    \caption{Comparison of the CPU cost as a function of the up quark mass (for $\beta=3.31$ ensembles). The comparison is biased by different (R)HMC parameters, number of nodes, and results such as the acceptance and correlation time which are not represented here, but it is clear that we do not hit any kind of ``Berlin wall'' singularity at zero PCAC mass. }
  \label{fig:timepertraj}
  \end{center}
  \end{figure}

As shown in  Figure~\ref{fig:timepertraj}, the CPU cost to generate those ensembles depends only mildly on $m_u^{\bare}$. A similar statement can be made on the condition number of the Dirac operator (as checked on a few configurations), which is again interpreted as an effect of the absence of light pion. 

\begin{figure}[thb]
  \begin{center}
    \hfill
    \includegraphics*[width=0.4 \textwidth,clip=true]{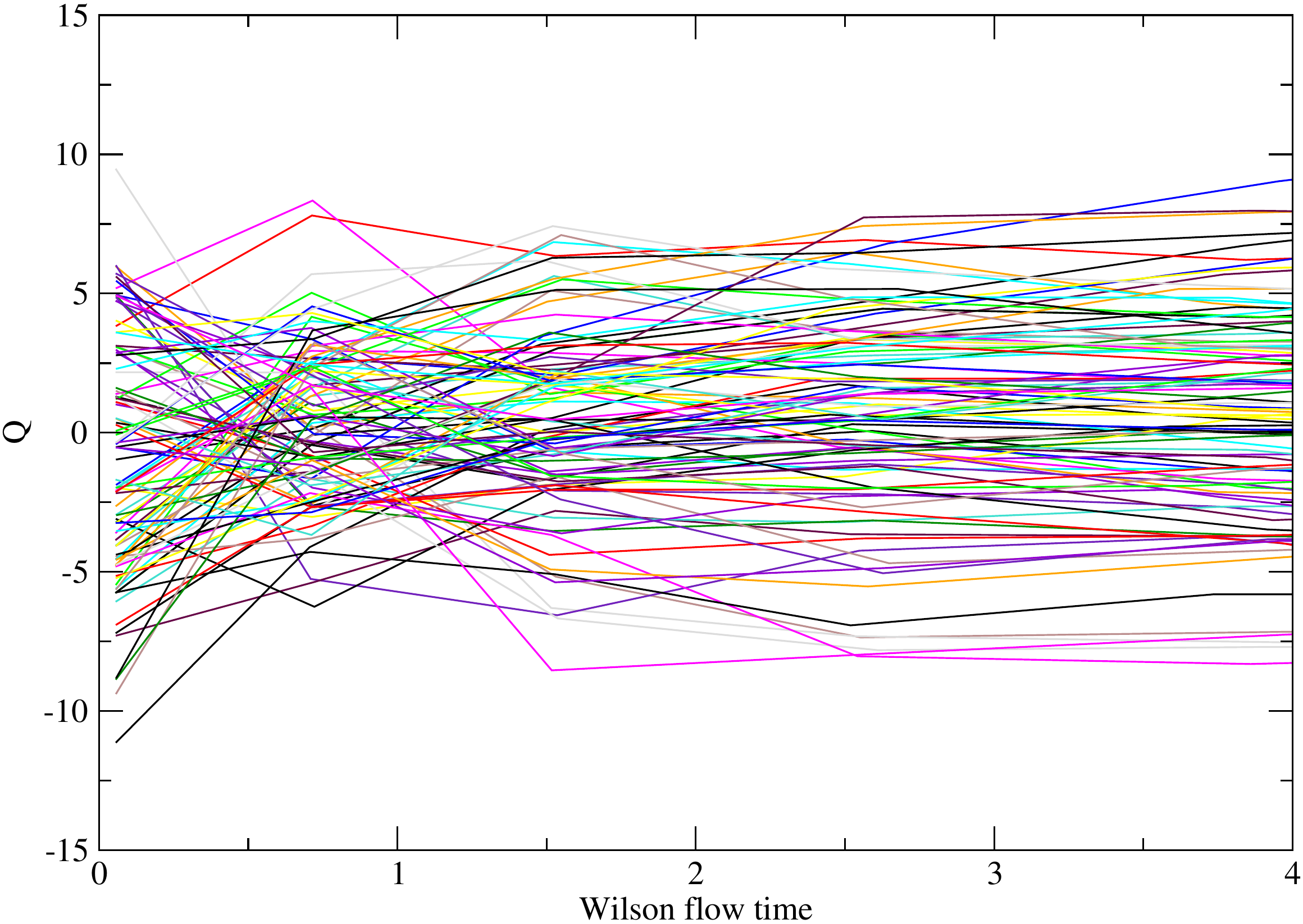}
    \hfill
    \includegraphics*[width=0.4 \textwidth,clip=true]{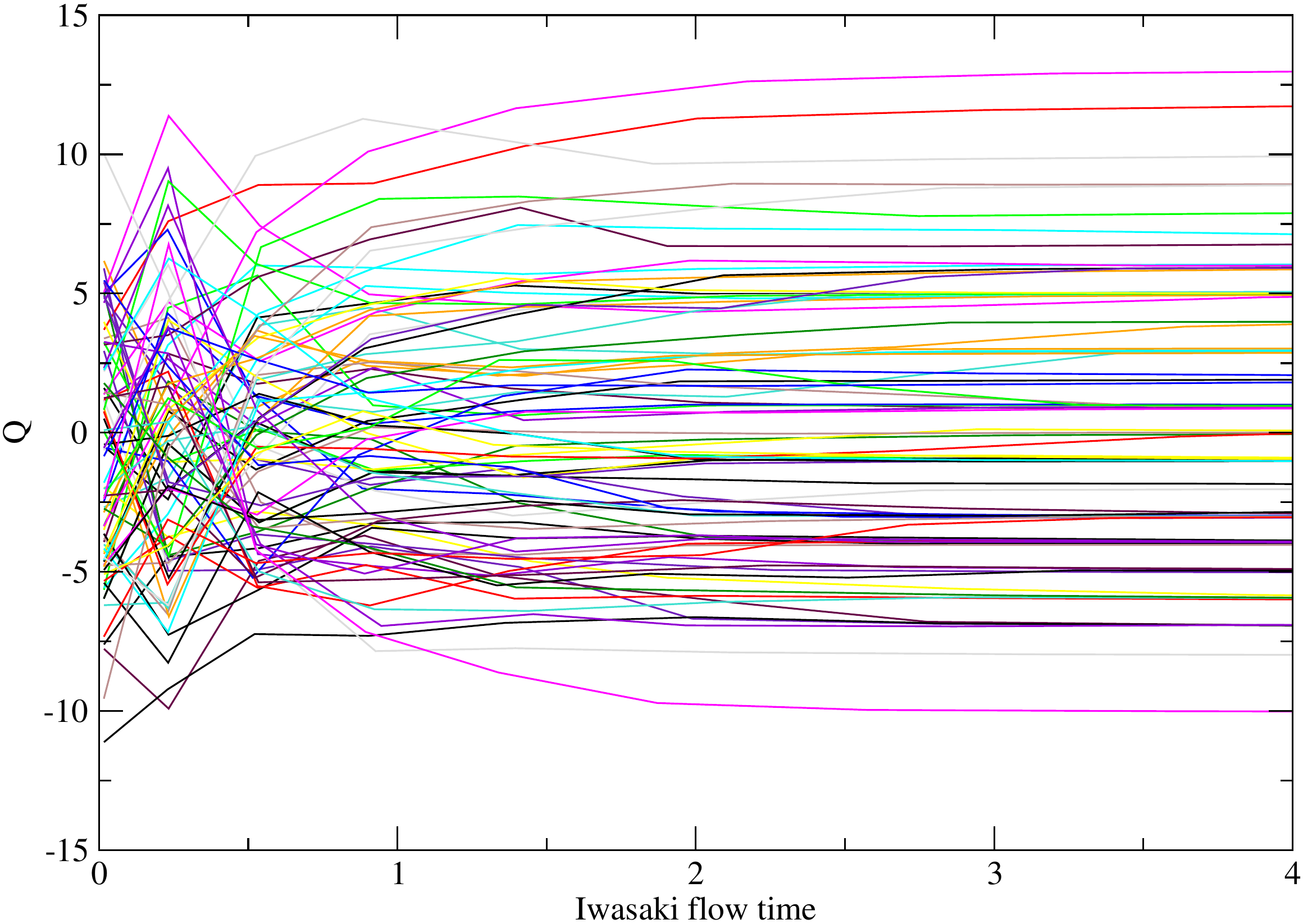}
    \hfill
    \\
    \vspace{0.6cm}
    \hfill
    \includegraphics*[width=0.4 \textwidth,clip=true]{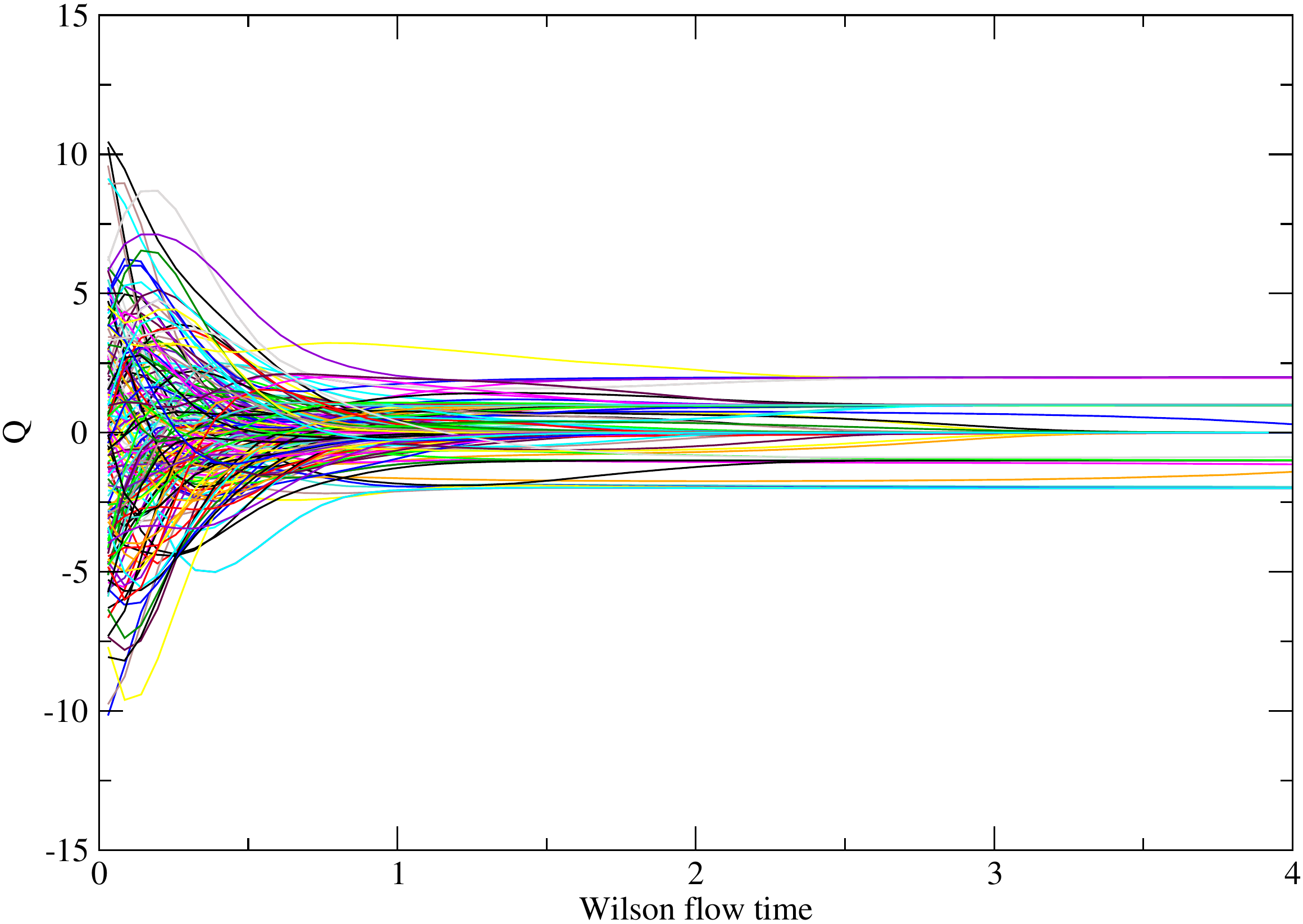}
    \hfill
    \includegraphics*[width=0.4 \textwidth,clip=true]{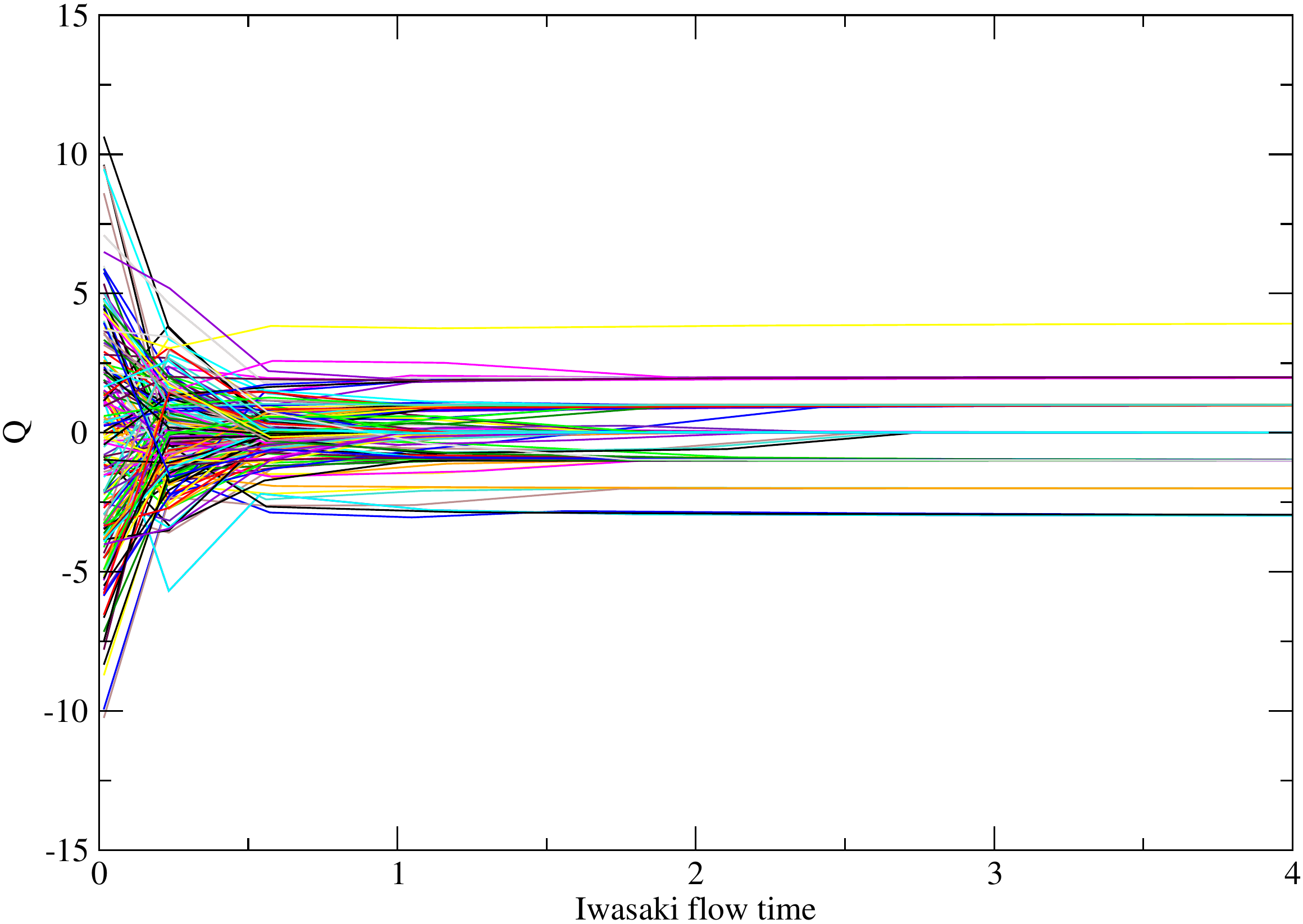}
    \hfill
    \caption{For coarse ensembles (top half, $\beta=3.31$) the choice of gradient flow action has a strong influence on the convergence and the stability of the asymptotic regime to integer values of $Q$. Actions with higher $c_1$ rectangle parameter (right side, Iwasaki) tend to give better results. The price to pay for this better behaviour is the possibility of an increase of discretisation errors and the survival of more unphysical dislocations which are not decoupled by the gradient flow. This typically leads to higher $\chi_t$ values for those actions (as seen in Figure~\ref{fig:chitw054}). At finer lattice spacings (bottom half, $\beta=3.7$), all actions tend to be more similar, qualitatively and quantitatively}
  \label{fig:tcharge}
  \end{center}
  \end{figure}

\begin{figure}[thb]
  \begin{center}
    \includegraphics[width=\linewidth]{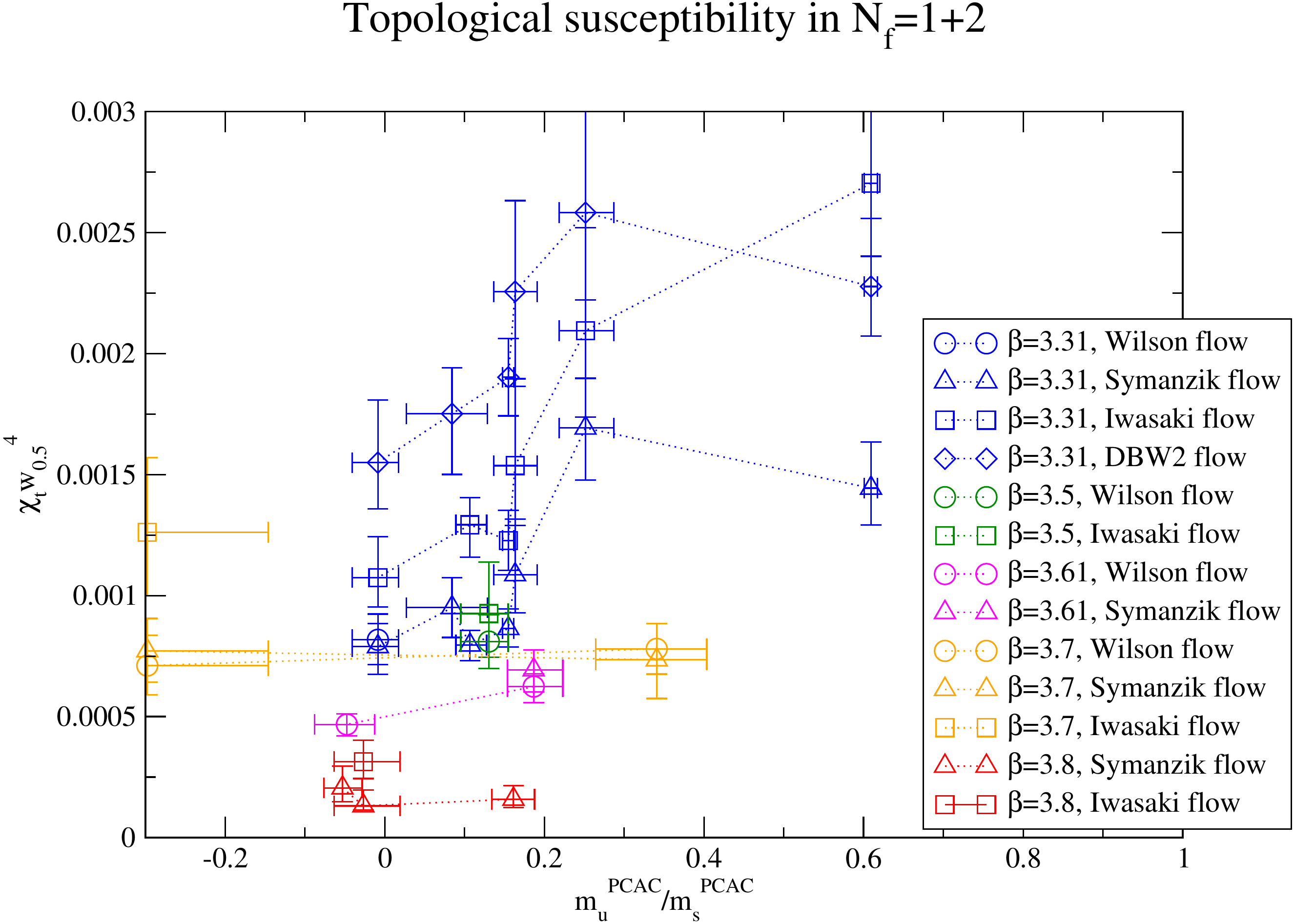}
    \caption{Smaller $m_u^{PCAC}$ values correspond to smaller $\chi_t$ as expected, but a large discretisation effect remains which prevents an exact cancellation of $\chi_t$ at finite lattice spacing. As lattice spacing is chosen to be smaller, this discretisation effect drops down and $\chi_t$ seems to reach zero within reasonable precision. However, the current data cannot exclude the possibility that the continuum limit of $\chi_t$ also vanishes at non-zero up mass, and the curvature/transition of the $\beta=3.31$ curves is still unexplained. Also, if we neglect this curvature and only consider their average slope, it tends to give a result much higher than what $N_f=2$ ChPT suggests. }
  \label{fig:chitw054}
  \end{center}
  \end{figure}

Finally, preliminary results are presented in Figures~\ref{fig:tcharge}~and~\ref{fig:chitw054}.

\begin{table}[thb]
\begin{center}
   \begin{tabular}{|c|c|c|c|c|}
     \hline
     $L^3\times T$ & $\beta$ & $m_u^{\bare}$ & $m_s^{\bare}$ & $N_{\rm conf}$\\
     \hline
     $16^3\times 32$ & $3.31$ & $-0.07$ & $-0.04$ & $151$ \\
     $16^3\times 32$ & $3.31$ & $-0.093$ & $-0.04$ & $172$ \\
     $16^3\times 32$ & $3.31$ & $-0.09756$ & $-0.04$ & $262$ \\
     $16^3\times 32$ & $3.31$ & $-0.1$ & $-0.04$ & $287$ \\
     $16^3\times 32$ & $3.31$ & $-0.102$ & $-0.04$ & $276$ \\
     $16^3\times 32$ & $3.31$ & $-0.105$ & $-0.04$ & $264$ \\
     $16^3\times 32$ & $3.31$ & $-0.108$ & $-0.04$ & $255$ \\
     $16^3\times 32$ & $3.31$ & $-0.11$ & $-0.04$ & $105$ \\
     $16^3\times 32$ & $3.31$ & $-0.12$ & $-0.04$ & $67$ \\
     \hline
     $16^3\times 32$ & $3.5$ & $-0.05$ & $-0.006$ & $180$ \\
     $16^3\times 32$ & $3.5$ & $-0.055$ & $-0.006$ & $252$ \\
     \hline
     $16^3\times 32$ & $3.61$ & $-0.03121$ & $+0.0045$ & $366$ \\
     $16^3\times 32$ & $3.61$ & $-0.0344$ & $+0.0045$ & $134$ \\
     $16^3\times 32$ & $3.61$ & $-0.0365$ & $+0.0045$ & $339$ \\
     $16^3\times 32$ & $3.61$ & $-0.04$ & $+0.0045$ & $548$ \\
     $16^3\times 32$ & $3.61$ & $-0.045$ & $+0.0045$ & $346$ \\
     \hline
     $16^3\times 32$ & $3.7$ & $-0.021$ & $0$ & $406$ \\
     $16^3\times 32$ & $3.7$ & $-0.027$ & $0$ & $274$ \\
     \hline
     $16^3\times 32$ & $3.8$ & $0$ & $0$ & $697$ \\
     $16^3\times 32$ & $3.8$ & $-0.021$ & $0$ & $434$ \\
     $16^3\times 32$ & $3.8$ & $-0.024$ & $0$ & $528$ \\
     $16^3\times 32$ & $3.8$ & $-0.024$ & $+0.02$ & $673$ \\
     $16^3\times 32$ & $3.8$ & $-0.03$ & $0$ & $356$ \\
     \hline
     \hline
     $24^3\times 48$ & $3.31$ & $-0.09756$ & $-0.04$ & $95$ \\
     $24^3\times 48$ & $3.31$ & $-0.1$ & $-0.04$ & $187$ \\
     $24^3\times 48$ & $3.31$ & $-0.108$ & $-0.04$ & $69$ \\
     \hline
     $24^3\times 48$ & $3.61$ & $-0.03121$ & $-0.04$ & $128$ \\
     $24^3\times 48$ & $3.61$ & $-0.0355$ & $-0.04$ & $291$ \\
     \hline
   \end{tabular}
   \caption{List of all ensembles available with statistics having reasonably reached thermalisation. Some of those ensembles have not (yet) been fully included in the analysis. $N_{\rm conf}$ includes non-thermalised configurations, and the number of trajectories between two configurations varies but is typically $5$. }
 \label{tab:ensembles}
\end{center}
\end{table}

\begin{table}[thb]
\begin{center}
   \begin{tabular}{|c|c|c|c|c|}
     \hline
     & \multicolumn{2}{c|}{$m_u^{PCAC}-m_u^{bare}$}
     & \multicolumn{2}{c|}{$M_{\pi^+} L$}\\
     \hline
     $\beta, m_u^{bare}$ & $16^3$ & $24^3$ & $16^3$ & $24^3$  \\
     \hline
     $3,31, -0.1$ & 0.1099(5) & 0.1094(7) & $4.90(8)$ & $8.30(2)$ \\
     \hline
     $3.61, -0.0365$ & 0.0355(7) & & $3.92(8)$ & \\
     $3.61, -0.0355$ & & 0.0369(9) & & $5.51(3)$ \\
     $3.61, -0.0344$ & 0.0362(11) & & $4.25(7)$ & \\
     \hline
   \end{tabular}
   \caption{Comparison of finite volume effects on the additive mass of the up quark. For the lightest masses and smallest volume we do see a bit of tension, but only $1.2\sigma$. $M_{\pi^+} L$ is given as an indication, but $\pi^0$ could be sensibly lighter. }
 \label{tab:m1624}
\end{center}
\end{table}

\begin{table}[thb]
\begin{center}
   \begin{tabular}{|c|c|c|}
     \hline
     & \multicolumn{2}{c|}{$\chi_t=\langle Q^2\rangle/V$ (lat. units)}\\
     \hline
     $\beta, m_u^{bare}$ (action) & $16^3$ & $24^3$ \\
     \hline
     $3.61, -0.0365$ (Wilson) & $0.80(7) \times10^{-5}$ &\\
     $3.61, -0.0355$ (Wilson) & & $1.20(14) \times10^{-5}$\\
     $3.61, -0.0344$ (Wilson) & $1.39(14) \times10^{-5}$ &\\
     \hline
   \end{tabular}
   \caption{Comparison of finite volume effects on $\chi_t$. If we linearly extrapolate $\chi_t$ between the $16^3$ data points, no significant finite volume effect can be seen. }
 \label{tab:chi1624}
\end{center}
\end{table}

\section{Summary}\label{sec:summary}

In order to gain some quantitative understanding of the additive renormalisation to $m_u$, we perform $N_f=1+2$ flavour QCD simulations. 
The preliminary results presented here contain some evidence of a suppression of $\chi_t$ by a massless quark, even with a single light flavour, when no light pion exists. 
On the other hand, our ability to generate ensembles at very light $m_u$ for a reasonable cost suggests that the absence of light pion does have some effect on the dynamics.
Even more surprising is our ability to generate ensembles at negative $m_u^{PCAC}$, without hitting any obvious sign problem. However, additional checks are to be performed, since the apparent success of the RHMC does not rigorously proves that the $\det D>0$ assumption was valid. Moreover, such a problem is expected anyway for $m_u^{PCAC}\approx -m_d^{PCAC}$, with a CP-violating transition to the Dashen phase, but it would not be directly connected to the strong CP problem. 


Our data shows an intriguing transition for a $10\ \MeV$-ish up quark (with unphysically large down quark), which could be related to the additive renormalisation. 
Studies with more statistics and a wide range of lattice parameters are on-going.\\

\section{Acknowledgement}\label{sec:acknow}

This work is in part based on Bridge++ code
(http://bridge.kek.jp/Lattice-code/). We thank Hideo Matsufuru for providing us with this code and for his support. 

This work is supported by JSPS KAKENHI Grant No. 15H03669, 15KK0176, 
MEXT KAKENHI Grant No. 25105011 (RK), the Large Scale Simulation Program 
No. 16/17-28 of High Energy Accelerator Research Organization (KEK), 
and Interdisciplinary Computational Science Program No. 17a15 in CCS, 
University of Tsukuba.

\clearpage
\bibliography{Lattice2017_275_FRISON}

\end{document}